\documentclass[12pt]{article}
\usepackage{pic04}
\usepackage{hyperref}
\usepackage{url}
\usepackage{graphicx}

\def\be{\begin{equation}}
\def\ee{\end{equation}}
\def\bea{\begin{eqnarray}}
\def\eea{\end{eqnarray}}
\def\ra{\rightarrow}

\def\dkpmn{D^+ \rightarrow K^-\pi^+\mu^+\nu}

\def\dkskpmn{D^+ \rightarrow \overline{K}^{*0}(K^-\pi^+)\mu^+\nu}
\def\dsphikkmn{D_s^+ \rightarrow \phi(K^-K^+)\mu^+\nu}
\def\costv{\cos \theta_V}
\def\dksmn{D^+ \rightarrow \overline{K}^{*0}\mu^+\nu}

\def\dsphikkp{D_s^+ \rightarrow \phi(K^-K^+)\pi^+}
\def\knp{K\,\pi}

\begin{document}

\title{\bf CHARM DECAYS}
\author{
Daniele Pedrini \\
{\em INFN Sezione di Milano, Via Celoria 16 20133 Milano, Italy}}
\maketitle

%
%


%
%
\vspace{4.5cm}


\baselineskip=14.5pt
\begin{abstract}
To write a review on charm decays is not an easy task because of the large amount (fortunately!) 
of experimental results in the recent years. I was forced to make a selection of topics. 
I am going to discuss with the eyes of an experimentalist: $D^0$-$\overline{D}^0$ mixing, 
\emph{CP} violation, rare decays, charm lifetimes, semileptonic decays, hadronic decays, 
and the big surprise of the $D_{sJ}$ states. Obviously there are some missing topics, however 
I hope to show clearly my point of view.
\end{abstract}
\newpage

\baselineskip=17pt

\section{Introduction}

Standard Model contributions to $D^0$-$\overline{D^0}$ mixing and \emph{CP} violation
in charm decays are greatly suppressed compared to similar contributions in the 
strange and bottom sectors. Investigations of the K and B systems have and will
continue to play a central role in our quest to understand flavor 
physics, but investigations of the charm-quark sector are 
fundamental too. Since charm is the only {\it up-type} quark for which 
the decay modes can be studied, it has a unique role to investigate flavor physics.
Charm allows a complementary probe of Standard Model beyond to that attainable from the 
{\it down-type} sector. It is indeed this suppression that makes searching for charm 
mixing and \emph{CP} violation interesting: it provides opportunities to search for 
New Physics beyond the Standard Model.

\section{$D^0$-$\overline{D}^0$ mixing}

Mixing occurs because the two weak eigenstates $D^0$ and $\overline{D}^0$ are not 
the mass eigenstates. If \emph{CP} is conserved, 
$\displaystyle D_{1,2}=\frac{D^0 \pm \overline{D}^0}{\sqrt{2}}$ 
are mass and \emph{CP} eigenstates with $\Delta \Gamma$ and $\Delta M$ different from 0.
Experimental limits on $D^0$-$\overline{D}^0$ mixing indicate that $\Delta \Gamma << \Gamma$
and $\Delta M << \Gamma$. It is then usual to define:

\begin{equation}
 x=\frac{\Delta M}{\Gamma}, \qquad y=\frac{\Delta \Gamma}{2\Gamma}
\end{equation}

 In the case of \emph{CP} conservation (and if $|x|, |y| \ll 1$) the amplitude for 
$D^0$ to mix into $\overline{D}^0$ assumes a very simple expression:

\begin{equation}
A_{mix}\sim \frac{y+ix}{2}\, \Gamma t\, e^{-\frac{\Gamma t}{2}}
\end{equation}

 To compute the transition rate, one has to take into account other processes that 
can interfere with mixing. In the case of the $D^0$ hadronic decays, the transition
$D^0 \to \overline{f}$ is fed both by mixing and Doubly Cabibbo Suppressed
Decay(DCSD):

\begin{eqnarray*}
\frac{dN_{D^0 \to \overline{f}}}{dt} \propto \left|A_{mix} + 
\sqrt{R_D} e^{-i\delta} e^{-\Gamma t/2}\right|^2 =
\end{eqnarray*} 
\begin{equation}
= \left[\left(\frac{x^2 + y^2}{2}\right)\,\frac{\Gamma^2 t^2}{2}+ \sqrt{R_D}(-x\sin\delta + y \cos \delta)\,
 \Gamma t + R_D\right] e^{-\Gamma t}
\label{equaws}
\end{equation} 

where $\displaystyle \delta$ is the strong phase shift due to Final State Interactions (FSI) 
between the DCSD ($\displaystyle D^0 \to \overline{f}$) and the Cabibbo Favored 
mode($\displaystyle \overline{D^0} \to \overline{f}$), and $\displaystyle R_D$ is their relative 
Branching Ratio. The net effect of $\displaystyle \delta$ is to rotate the definition of 
$x$ and $y$ by the same angle, that is  $\displaystyle x^\prime =(x\cos\delta + y\sin\delta)$ 
and $\displaystyle y^\prime =(-x\sin\delta + y\cos\delta)$.

In contrast with $D^0$ hadronic decays, in the $D^0$ semileptonic decays the DCSD terms
are absents and the entire mixing rate is proportional to the square of $x$ and $y$
multiplied by $t^2$. 

  A wide range (roughly 7 orders of magnitude) of Standard Model and 
non-Standard Model predictions appear in literature\cite{Nelson,Petrov}. Most of 
these theoretical predictions are well below the one percent level for both $x$ and $y$.
Very recently, however, a theoretical prediction has been published\cite{Falk} that shows
it is possible for the Standard Model mixing mechanisms to generate $x$ and $y$ values 
of about one percent, complicating searches for new physics. 

 A well known method to measure mixing parameters is to form the ratio of the number of 
Wrong Sign (WS) decays $\displaystyle D^0 \to K^+ \mu^-\overline{\nu}_\mu$ to the Right Sign (RS) 
$\displaystyle D^0 \to K^- \mu^+\nu_\mu$ to measure directly $ \displaystyle R_{mix}$
(throughout this paper the charge conjugate state is implied, unless otherwise noted). 
The limiting factor for this method is the background under the WS signal.

 Another possibility would be to measure directly $x$ or $y$. The measurement of $x$ is not feasible
since it would require a mass resolution of about 25 $\mu$eV in the most favorable case 
(assuming the mixing entirely due to $x$). On the other hand, the measurement of $y$ turns out to be 
one of the best experimental probes of charm mixing. Experimentally instead of $\displaystyle y$ one 
measures $\displaystyle y_{CP}$ which, in the limit of \emph{CP}-symmetry(see section on \emph{CP}
violation for present limits), equals $\displaystyle y$. 
The easiest way to measure $\displaystyle y_{CP}$ is by comparing the lifetimes of $D^0 \to K^-K^+$ and 
$D^0 \to K^-\pi^+$. In fact, assuming that the $K^-\pi^+$ final state is an equal mixture of 
\emph{CP}-even and \emph{CP}-odd states, one finds that: 

\begin{equation}
 y_{CP}=\frac{\Gamma(CP~\mathrm{even})-\Gamma(CP~\mathrm{odd})}
 {\Gamma(CP~\mathrm{even})+\Gamma(CP~\mathrm{odd})}=
 \frac{\tau(D^0 \to K^-\pi^+)}{\tau(D^0 \to K^-K^+)}-1
\end{equation}

 The last method is based on the measurement of the interference term between the mixing
and DCSD amplitudes (see equation~\ref{equaws}): 
$\sqrt{R_D}(-x\sin\delta + y \cos \delta)\, \Gamma t \, e^{-\Gamma t}$.

 Information on mixing is massively improving due to the advent of new very high statistics data
from BABAR and BELLE, both through the study of the interference of mixing with DCSD and lifetime 
differences. The current best determination of $y_{CP}$ is $0.008 \pm 0.004~^{+0.005}_{-0.004}$\cite{BABAR_ycp}
and it will be very interesting to see if mixing does occur at the percent level.
 
 There is an excellent review of this subject (and of \emph{CP} violation and rare decays too) by 
 Burdman and Shipsey\cite{Shipsey} and I suggest the interested reader to look at this review for 
 more details.

\section{\emph{CP} violation}
 \emph{CP} violation occurs if the decay rate for a particle differs from the
decay rate of its \emph{CP}-conjugate particle\cite{BigiSanda}. \emph{CP} violation, 
which in the Standard Model (SM) is a consequence of a complex amplitude in the 
Cabibbo-Kobayashi-Maskawa (CKM) matrix, has been observed in $K$ and $B$
decays. In charm meson decays (as well as in $K$ and $B$) two classes of 
\emph{CP} violation exist: indirect and direct. In the case of direct violation, 
\emph{CP} violating effects occur in a decay 
process only if the decay amplitude is the sum of two different parts\cite{Buccella}, 
whose phases are made of a weak (CKM) and a strong contribution due to FSI: 
$\quad \alpha = A e^{i\delta_1} + B e^{i\delta_2}$.

 The weak contributions to the phases
change sign when going to the \emph{CP}-conjugate process, 
while the strong ones do not. Therefore \emph{CP} violating
asymmetry will be:

\begin{equation}
 A_{CP} = \frac{|\alpha|^2 - |\overline{\alpha}|^2}{|\alpha|^2 + |\overline{\alpha}|^2} =
  \frac{2Im(AB^\star)sin(\delta_2 - \delta_1)}{|A|^2+|B|^2 + 2Re(AB^\star)cos(\delta_2 - \delta_1)} 
\end{equation}

 In singly Cabibbo-suppressed $D$ decays, penguin terms 
in the effective Hamiltonian may provide the different phases of the 
two weak amplitudes.

 Compared to the strange and bottom sectors, the SM predictions of \emph{CP} 
violation for charm decays are much smaller \cite{Shipsey,Buccella,Golden,Close}, 
making the charm sector a good place to test the SM and to 
look for evidence of new physics. In the SM, direct \emph{CP} 
violating asymmetries in $D$ decays are predicted to be largest in
singly Cabibbo-suppressed decays, at most $10^{-3}$, and non-existent in 
Cabibbo-favored and doubly Cabibbo-suppressed decays. However, a \emph{CP} 
asymmetry could occur in the decay modes $D \to K_s \rm{n}\pi$ due to interference
between Cabibbo-favored and doubly Cabibbo-suppressed decays\cite{BigiSanda}.

 Before searching for a \emph{CP} asymmetry one must account for
differences, at the production level, between $D$ and $\overline{D}$ 
in photoproduction and hadroproduction fixed target experiments(the 
hadronization process, in the presence of remnant quarks from the nucleon, 
gives rise to production asymmetries). This is done usually 
by normalizing to the Cabibbo-favored modes, with the additional benefit that 
most of the corrections due to inefficiencies cancel out, reducing systematic 
uncertainties. An implicit assumption is that there is no measurable \emph{CP} 
violation in the Cabibbo-favored decays.

 The \emph{CP} asymmetry parameter measures the direct \emph{CP} 
asymmetry in the case of $D^+$ and the combined direct and indirect 
\emph{CP} asymmetries for $D^0$.
  
\begin{table}
\centering
\caption{\it \emph{CP} asymmetry measurements.}
\begin{tabular}{|l|c|c|} \hline
 Decay mode &  &  \\
\hline
\hline
                          & $\mathbf{E791}$               & $\mathbf{CLEO}$              \\
$D^0 \to K^-K^+$          & $ -0.010 \pm 0.049 \pm 0.012$ & $+0.000 \pm 0.022 \pm 0.008$ \\
$D^0 \to \pi^-\pi^+$      & $-0.049 \pm 0.078 \pm 0.030$  & $+0.030 \pm 0.032 \pm 0.008$ \\
$D^0 \to K_S K_S$         &                               & $-0.23 \pm 0.19$             \\ 
$D^0 \to K_S\pi^0$        &                               & $+0.001 \pm 0.013$           \\
$D^0 \to \pi^0\pi^0$      &                               & $+0.001 \pm 0.048$           \\
$D^+ \to K^-K^+\pi^+$     & $-0.014 \pm 0.029$            &                              \\
$D^+ \to \pi^-\pi^+\pi^+$ & $-0.017 \pm 0.042 $           &                              \\

                      & $\mathbf{FOCUS}$              & $\mathbf{CDF}$                   \\			  
$D^0 \to K^-K^+$      & $-0.001 \pm 0.022 \pm 0.015$  & $0.020 \pm 0.012 \pm 0.006$      \\				  
$D^0 \to \pi^-\pi^+$  & $+0.048 \pm 0.039 \pm 0.025$  & $0.030 \pm 0.013 \pm 0.006$      \\
$D^+ \to K^-K^+\pi^+$ & $+0.006 \pm 0.011 \pm 0.005$  &                                  \\
$D^+ \to K_S\pi^+$    & $ -0.016 \pm 0.015 \pm 0.009$ &                                  \\
$D^+ \to K_SK^+$      & $ +0.071 \pm 0.061 \pm 0.012$ &                                  \\
			  
\hline 
\end{tabular}
\label{CPtable}
\end{table}
 
 Table \ref{CPtable} summarizes the current standing of the \emph{CP} 
asymmetry measurements. From this summary I can draw some conclusions:
 
\begin{itemize}
\item the $1$\% level have been reached for some decay modes, that is still an order
 of magnitude bigger than the theoretical expected values, but represents a substantial
 improvement
\item the measured \emph{CP} asymmetries are all consistent with zero within the errors
\item there is no evidence of \emph{CP} violation in the charm sector
\end{itemize}

In my opinion, however, with large data sample of reconstructed charm particles
it will be possible to study \emph{CP} violation (or better \emph{T} violation)
using the \emph{T}-odd correlation obtained by a triple product. This type of 
measurement in the charm sector was suggested by Bigi sometime ago\cite{Bigitriple}.
For example, consider the decay mode $D^0 \rightarrow K^-K^+\pi^-\pi^+$ where one can 
form a \emph{T}-odd correlation with the momenta:

\begin{equation}
 C_T = < \vec{p}_{K^+} \cdot ( \vec{p}_{\pi^+} \times \vec{p}_{\pi^-} ) >
\end{equation}
 
 under time reversal $C_T \rightarrow -C_T$ hence the name \emph{T}-odd.
Since time reversal is implemented by an anti-unitary operator, $C_T \ne 0$ can be induced
by FSI. So $C_T \ne 0$ does not establish \emph{T} violation
because FSI can act as an impostor here. This ambiguity can be resolved by computing
$\overline{C_T}$ from $\overline{D^0} \rightarrow K^-K^+\pi^-\pi^+$:
 
\begin{equation}
 \overline{C_T} = < \vec{p}_{K^-} \cdot ( \vec{p}_{\pi^-} \times \vec{p}_{\pi^+} ) >
\end{equation}

Finding  $C_T \ne -\overline{C_T}$ establishes \emph{CP} violation without further troubles.
FOCUS tried to make this measurement but the statistics was low and found 
no evidence for \emph{T} violation. My advice to present and future experiments is to pursue
this type of search. It is indeed a clean way to search for \emph{CP} violation. 

Another interesting way to search for \emph{CP} violation is by means of the Dalitz plot analysis, this
will be discussed in the hadronic decay section.

\section{Rare decays}
One interesting way to search for physics beyond the Standard Model
is to look at decay modes that are extremely {\it rare} or {\it forbidden}
in the charm sector. In the charm sector the rare and forbidden decay modes can be 
split mainly into three categories:

\begin{itemize}

  \item[1)] Flavor Changing Neutral Current(FCNC) such as 
    $D^0 \to \ell^+ \ell^-$ and \\
    $D^+ \to h^+\ell^+\ell^-$

  \item[2)] Lepton Family Number Violating(LFNV) such as
   $D^+ \to h^+ \ell_1^+ \ell_2^-$
		
  \item[3)] Lepton Number Violating(LNV) such as 
   $D^+ \to h^- \ell_1^+ \ell_{1,2}^+$
\end{itemize}
where $h$ stands for $\pi,K$ and $\ell$ for $e,\mu$. 

The first decay modes(FCNC) are {\it rare}, where rare decays
usually means a process which proceeds via an internal quark loop in the 
Standard Model (forbidden at the tree level); i.e. cannot proceed via a 
single charged current W-exchange\cite{Schwartz}. The other two decay 
modes(LFNV and LNV) are strictly {\it forbidden} in the Standard Model. 

The FCNC decay mode $D^0 \to \ell^+ \ell^-$ can proceed via a W box diagram, 
also contributing to the $D^0-\overline{D^0}$ mixing. 
However it is expected to be helicity 
suppressed by a factor $(\frac{m_l}{m_D})^2$ and current 
estimates\cite{Schwartz,Hewett} for the branching fractions 
are of the order of $\sim 10^{-19}$. Although long-distance effects are predicted
to enhance these rates by several orders of magnitude, these expected
rates are out of the sensitivity of the present and near future experiments.
This is the reason why these decays modes are so attractive: any detection, 
in fact, will be a clear sign of physics beyond the Standard Model.
 
The predictions for the other FCNC decay modes, $D^+ \to h^+\ell^+\ell^-$, are
considerably larger. These decay modes can proceed via several penguin 
diagrams\cite{Schwartz}, but the 3-body decays do not suffer from helicity
suppression. The branching fraction estimates for these modes are of the order 
of $\sim 10^{-9}$.
In addition to these short distance diagrams there are contributions from 
long distance effects, such as photon pole amplitude or resonance 
decays\footnote{For example the final state $D^+ \to \pi^+\mu^+\mu^-$ can be
obtained via the decay chain $D^+ \to \phi\pi^+$ with $\phi \to \mu^+\mu^-$.
This mode has a composite branching fraction of $\sim 10^{-6}$, three order of
magnitude bigger than the corresponding non-resonant decay mode.However this
contribution can be handle properly excluding the kinematic regions of the
muon couple corresponding to the $\phi$ mass}, which can be much larger. 
The difficulty in the calculation of the long distance contributions is linked
to the hadronic uncertainties\cite{Hewett}. The estimates for these 
contributions are around $\sim 10^{-7}$; in all cases the long distance
contributions overwhelm those from the Standard Model short distance physics.

The LFNV and LNV decay modes test the conservation of the lepton family
and of the lepton number. There is no reason why the lepton number should 
be conserved, that is no fundamental principle analogous to the gauge 
invariance exists\cite{Schwartz}. Therefore lepton number conservation 
could break down, but the question is at what energy scale ? 
For example one can consider the decay mode $D^+ \to \pi^+\mu^+e^-$
where the particle H, supposed to mediate this interaction, 
contains the quantum number of quarks and leptons. It is thus 
referred as {\it leptoquark}\cite{Schwartz}.
Considering the Standard Model analogous process 
$D^+ \to \overline{K^0}\mu^+\nu$ (mediated by the $W^+$) and assuming the same 
coupling constant one can calculate that for an experimental sensitivity of 
$\sim 10^{-5}$ the non-observation of this mode implies a mass for the H 
particle greater than $800~ GeV/c^2$, or in other words one is probing the 
$800~ GeV/c^2$ mass scale.

Recently CDF published an upper limit for the decay $D^0 \to \mu^+ \mu^-$\cite{CDFmumu}
improving by a factor 2 the previous PDG 2002~\cite{PDG2002} limit. The search for New
Physic effects in charm rare decays is going on; the gap between the theoretical predictions 
and the experimental upper limits is still large, even if it is reducing in the recent years. 
There is already an upper limit ($D^+ \to \pi^+\mu^+\mu^-$) that tends to exclude a prediction of 
the Minimal Supersymmetric Standard Model\cite{MSSM}.

\section{Charm lifetimes}

\begin{figure*}[ht!]
\hbox to\hsize{\hss
\includegraphics[width=0.5\hsize]{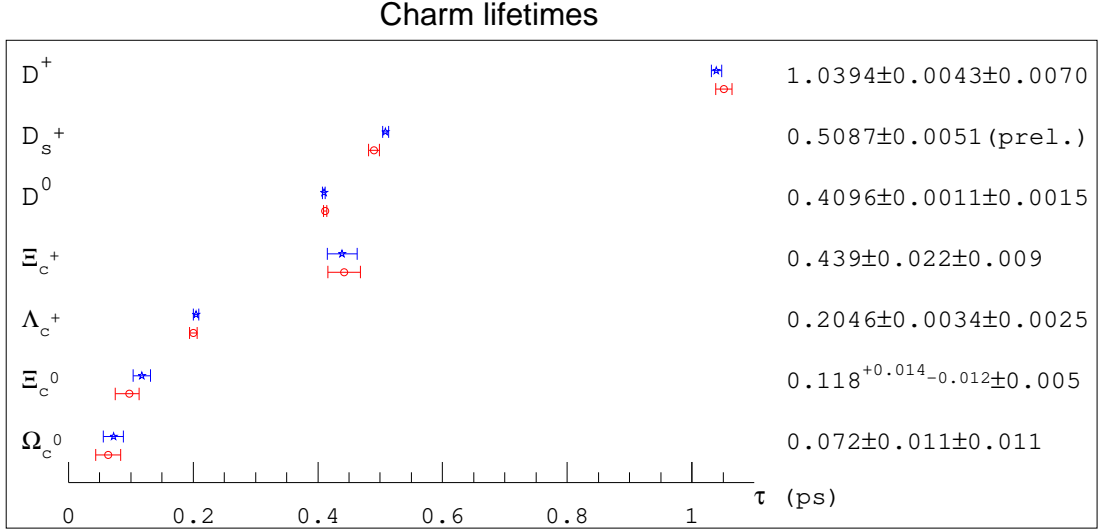}
\hss}
\caption{\it Charm particle lifetimes, comparison between the FOCUS lifetime 
measurements and the PDG 2002 values. The $\star$ are the FOCUS results reported 
also on the right, while the $\circ$ correspond to the PDG 2002 values. The PDG 2002 
values for $\Xi_c^+$ and $\Lambda_c$ include already the FOCUS measurements. 
\label{fig:lifetime}}
\end{figure*}

 The determination of lifetimes allows to convert the branching ratios measured by experiments
to partial decay rates predicted by theory. FOCUS is the only experiment (with the 
predecessor experiment E687) to have measured the lifetimes of all the weakly decaying charmed 
particles. This is particularly important when one forms the ratio of lifetimes because 
most of the systematic errors cancel out. Fig~\ref{fig:lifetime} shows a comparison between 
the PDG 2002~\cite{PDG2002} values and the FOCUS lifetime measurements. FOCUS produced new lifetimes 
results with precision better than the previous world average. An accurate measurement of the 
$D^0$ lifetime for the golden decay mode into $K\pi$ is a crucial ingredient to determine the 
lifetime difference, and consequently the parameter $y$ of the $D^0-\overline{D^0}$ mixing. 

 The increasingly precise measurements of the heavy quark lifetimes have
stimulated the further development of theoretical models, like the Heavy Quark Theory~\cite{HQE}, 
which are able to predict successfully the rich pattern of charm hadron lifetimes, that span 
one order of magnitude from the longest lived $D^+$ to the shortest lived $\Omega_c^0$. For the 
charm mesons a clear lifetime pattern emerges in agreement with the theoretical predictions:
$\tau(D^0)<\tau(D_s^+)<\tau(D^+)$.

 Even the expectations~\cite{HQE,Guberina,Voloshin} for the charm baryon lifetimes reproduce 
the data, which is quite remarkable since, in addition to the exchange diagram, there are 
constructive as well as destructive contributions to the decay rate. The experimental 
results lead to the following baryon lifetime hierarchy:
$\tau(\Omega_{c}^0)\leq\tau(\Xi_{c}^{0})<\tau(\Lambda_{c}^+)<\tau(\Xi_{c}^{+})$.

\section{Semileptonic Decays}

The semileptonic decays of heavy flavored particles can be calculated from first principles.
Involving a lepton in the final decay stage implies that one does not have to worry about the usual
final state interaction between hadrons.  The possible complications coming
from QCD corrections of the decay process are contained in form factors.
The form factors can be calculated by various models: HQET, Lattice Gauge Theories 
and Quark models. The angular distributions and invariant masses among the decay
products would determine the form factors ratios while the branching
ratio measurements and information from the CKM matrix would give the absolute
scale for the form factors. Charm semileptonic decay provide a high quality lattice
calibration crucial to reduce future systematic error in the Unitarity Triangle. The same 
techniques validated in charm can be applied to beauty.

 There are recent results from CLEO\cite{CLEO04} on the decays 
$D \to \mathit{Pseudoscalar} ~\ell ~\nu$
both for the Cabibbo favored $D^0 \to K^- e^+\nu$ and for the Cabibbo suppressed 
$D^0 \to \pi^- e^+\nu$. These measurements turned out in a big advance in precision 
for the two branching ratios measurements.  
   
For the last 20 years people regarded the $\dkpmn$  decays (that is $D \to \mathit{Vector} ~\ell ~\nu$)
as 100\% $\dkskpmn$ events. This situation changed when FOCUS started to analyze $\dkpmn$ decays 
to get form factors. During this form factor analysis\cite{focus_sw}, FOCUS checked the angular 
distribution of Kaon in the $\knp$ rest frame ($\costv)$ and found that it showed a huge 
forward-backward asymmetry below the $K^*(892)$ pole mass while almost no asymmetry
above the pole.  Since the $K^*$ is a P-wave, pure $K^* \ra K\pi$ decays would have shown only a 
symmetric forward-backward $\costv$ distribution over the entire $\knp$ invariant mass range. This 
suggests a possible quantum mechanics interference effect. A simple approach to emulate the interference
effect is adding a spin zero amplitude in the matrix elements of the $\dkpmn$ decays. FOCUS tried 
a constant amplitude with a phase, $A \exp(i\delta)$, in the place where the $K^*$ couples to the 
spin zero component of the $W^+$ particle. FOCUS determined that the pure $\dksmn$ events are 
94.5\% of the selected events and therefore the BR has to be modified accordingly.

FOCUS measured also the relative branching ratio between $\dsphikkmn$
and $\dsphikkp$ decays. This measurement is comparable with all the other measurements 
in this channel and there is no evidence for s-wave interference in $\dsphikkmn$.

For what concerns the $D \to Vector ~\ell ~\nu$ form factors, the vector and axial form factors are
generally parametrized by a pole dominance form, and the experiments usually measures 
$r_V\equiv V(0)/A_1(0)$ and $r_2\equiv A_2(0)/A_1(0)$.
For the decay mode $\dksmn$ the experimental results are getting very precise and more theoretical
calculations are needed. Theoretically the $\dsphikkmn$ form factors should be within 10\% of $\dksmn$.
The measured $r_V$ values were consistent, but $r_2$ was a factor 2 higher. Anyway the very recent
FOCUS measurement\cite{focus_ff} has consistent $r_2$ value as well.

I want to conclude this section with a question: will there be similar effects (interference) 
in other charm (or beauty) semileptonic channels? We will see, in the meanwhile the 
analyses of other semileptonic charm decay modes are actively going on and we expect new results soon.

\section{Hadronic Decays} 

If we have found complications in the simpler semileptonic decays, one can easily imagine that
the hadronic decays are much more complicated. 

 It turned out that FSI play a central role in the hadronic decays. For example, the FOCUS recent 
analysis\cite{focus_kkpipi} of the branching ratio 
$\Gamma(D^{0} \rightarrow K^-K^+)/\Gamma(D^{0}\rightarrow \pi^-\pi^+)$ (known as a
long standing puzzle of charm decays) confirm that FSI are fundamental. Actually an isospin analysis 
of the channels $D\rightarrow KK$ and $D\rightarrow \pi \pi $ reveals that the elastic FSI cannot 
account for all the large deviation from unity of this ratio($\sim 2.81$). The most reasonable 
explanation seems to be the inelastic FSI that also allow for the transition $KK \rightarrow \pi\pi$.

 For the multi-body modes, where resonances are present, the amplitude analysis 
(Dalitz plot analysis) is the correct way to determine the resonant substructure of the decays.

What does one learn from Dalitz plot analysis? Bands indicate resonance contribution. For spinless
parents, like $D^+$ or $D_s^+$, the number of nodes in the bands give the resonance spin. Interference 
pattern gives relative phases and amplitudes\cite{Wiss_varenna}, so one has access to a complete set of
information not only the branching ratio!

But there is a complication for charm Dalitz plot analysis: one needs to face the problem of dealing 
with light scalar particles populating charm meson hadronic decays, such as $D \to \pi\pi\pi$ and
$D \to K\pi\pi$. This requires the understanding of light-quark hadronic physics, including the 
riddle of $\sigma (600)$ and $\kappa (900)$ (that is $\pi\pi$ and $K\pi$ states produced close to threshold)
whose existence is still controversial.

Let me consider the decay $D \to r(\to 1 + 2) + 3$, the problem is to write the propagator for the 
resonance r. For a well-defined wave with specific isospin and spin (IJ) characterized by narrow and 
well-isolated resonances, the propagator can be approximated by a single Breit-Wigner. In the so-called
isobar model the Dalitz plot amplitude is given by the following sum:

\begin{equation}
 M =  a_0 e^{i\delta_0} + \sum_{j} a_j~e^{i\delta_j}A_j
\end{equation}
  
where the first term is for the non-resonant component and in the sum the index j runs 
over the resonances. The  fit parameters are $a_i$ and $\delta_i$, and $A_j$ is
$A_j = F_D F_r \times |\vec{p_1}|^j|\vec{p_3}|^j P_j(\cos\theta^r_{13})\times BW(m_{12}^2)$
(see \cite{e687_kkpi} for more details). Nearly all charm Dalitz plot analysis use 
the isobar model. In contrast, when the specific IJ-wave is characterized by large and heavily 
overlapping resonances (just as the scalars), to write the propagator is not so simple. Indeed,
it is very easy to realize that the propagation is no longer dominated by a single resonance but
it is the result of a complicated interplay among resonances. In this case, it can be demonstrated
on very general ground that the propagator may be written in the context of the K-matrix 
approach\cite{Wigner,Chung} as $(I - iK \cdot \rho)^{-1}$, where $K$ is the matrix for 
the scattering of particles $1$ and $2$, that is to write down the propagator one needs the 
scattering matrix. 

What is the problem with the isobar model? The problem is that the sum of two (or more) Breit-Wigner
for broad and overlapping resonances does not respect unitary. Moreover the representation in the 
Argand plot shows that the phase is not properly evaluated. On the contrary adding two K matrices 
respects unitary and has the correct behavior in the Argand plot.

Very recently FOCUS published a pioneer work\cite{focus_kmatrix} using, for the first time in the 
analyses of charm decays (applied to the decay modes $D_s^+(D^+) \to \pi^-\pi^+\pi^+$), the 
formalism of K-matrix. The FOCUS amplitude was written as a sum:

\begin{equation}
 M = a_0 e^{i\delta_0} + F + \sum_{j}^{J>0} a_j~e^{i\delta_j}A_j
\end{equation}

where the first term is the usual non-resonant term, the second term (F) models the S-wave (the scalars),
and the last term is an isobar Breit-Wigner sum for higher spin resonances.
FOCUS obtains reasonable fits with no re-tuning of the K-matrix parameters\cite{Anisovich} 
(this was not obvious at the beginning of the analysis) and no need to invoke new 
resonances (such as $\sigma(600)$).

With Dalitz plot analysis we have a new tool to search for \emph{CP} violation because one gets amplitude
coefficients and phases. There are two recent results of this type of search. The first one is from the 
Dalitz plot analysis of the decay mode $D^0 \to K_s \pi^+\pi^-$ by CLEO\cite{CLEO_cp}. 
They used a smart way to represent the CP violating parameters. They showed upper limits and computed
also an integrated (over the Dalitz) asymmetry. The second result was shown by FOCUS in 
2002\cite{focus_ichep02}. FOCUS presented a Dalitz plot analysis of the CSD $D^+ \to K^-K^+\pi^-$, 
determining the phases and the amplitude coefficients separately for $D^+$ and $D^-$. The fact that
this analysis did not produce a paper should be clear from the previous discussion: a K-matrix analysis 
is necessary to handle properly the scalar resonances present in this final state. The intention of FOCUS 
is to revisit this analysis and publish soon a paper. 

\section{$D_{sJ}$ states}

It was one of the hot topic of the year 2003. I want just briefly to remind that in spring 2003 
BABAR discovered\cite{BABAR_dsj} a new particle $D_s(2317)$ which decays in $D_s^+\pi^0$ violating
isospin. This discovery was later confirmed by CLEO and BELLE. Then
a second state $D_{sJ}(2463)$ was found by CLEO\cite{CLEO_dsj} which decays in $D_s^*\pi^0$, and 
later confirmed by BABAR and BELLE. Strange property of these states 
is their surprisingly low mass compared to the potential model expectations. Their mass are practically 
equal to those of similar states in $c \overline{u}$ system: $D_0^{*0}(\sim 2308)$ and 
$D_1^{\prime 0}(\sim 2427)$. The $J^P$ of $D_s(2317)$ seems to be $0^+$, this is suggested by the 
low mass, absence of the decay $D_s^+\gamma$ and absence of the decay $D_s^+\pi^+\pi^-$. For what concerns 
$D_{sJ}(2463)$ it seems to have unnatural spin-parity $1^+$, as suggested by the analysis of 
BELLE\cite{BELLE_dsj}. In addition the widths of these two states ($\Gamma < 10 MeV$) are consistent with 
the experimental resolution.

Looking at the spectroscopy of the $c \overline{s}$ states one can conclude that the new states do not fit 
well: mass below the $DK~[D^*K]$ threshold. If interpreted as ordinary $c \overline{s}$ states, they decay 
mainly by isospin-violating $\pi$ emission thus having widths quite narrow. A possible decay mechanism could 
be through a virtual $\eta$ followed by $\eta-\pi^0$ mixing. It is interesting to note that
$M(D_s(2317))-M(D_s(1969) \cong M(D_{sJ}(2463))-M(D^*_s(2112))$ as predicted by models based on HQET and
chiral symmetry\cite{Bardeen}. A lot of interested by theorists: more than 40 papers have been already 
published.

\section{Conclusions}

 At 30 years from the discovery of the {\it c} quark the physics analyses of the first heavy quark
have reached a complete maturity. With the large statistics now available in the charm sector we start
to see unexpected effects which complicate the interpretation of the decay processes, both in semileptonic
and hadronic decays. For sure we know that FSI play a crucial role and that the physics of the light hadrons
is important for the correct interpretation of the charm decays. This means lessons for the $B$ decays? 
I think yes.

Exciting new states $D_{sJ}$ have been found. They maintain the promise of their constituent quarks to be
at the same time charm and strange.

Finally let me remind the players in this sector. While E791 and FOCUS are finishing their analysis, the present
and the near future are dominated by the results of BABAR, BELLE, CLEO-c and unexpectedly CDF. In the long range
I foresee interesting results in this sector by BTeV and LHC-b.

\end{document}